\documentclass[preprint]{aastex62}
\usepackage{amsmath,amssymb}
\usepackage{textcomp}
\usepackage{booktabs}
\usepackage{bm}

\newcommand\ltsima{$\; \buildrel <\over\sim \;$}
\newcommand\simlt{\lower.5ex\hbox{\ltsima}}
\newcommand\gtsima{$\; \buildrel >\over\sim \;$}
\newcommand\simgt{\lower.5ex\hbox{\gtsima}}

\shorttitle{Galactic Distribution of Planets}
\shortauthors{Koshimoto et al.}

\begin{document}

\title{No Large Dependence of Planet Frequency on Galactocentric Distance}

\author{Naoki Koshimoto}
\affil{Laboratory for Exoplanets and Stellar Astrophysics, NASA/Goddard Space Flight Center, Greenbelt, MD 20771, USA}
\affil{Department of Astronomy, University of Maryland, College Park, MD 20742, USA}
\affil{Department of Earth and Space Science, Graduate School of Science, Osaka University, Toyonaka, Osaka 560-0043, Japan}

\author{David P. Bennett}
\affil{Laboratory for Exoplanets and Stellar Astrophysics, NASA/Goddard Space Flight Center, Greenbelt, MD 20771, USA}
\affil{Department of Astronomy, University of Maryland, College Park, MD 20742, USA}

\author{Daisuke Suzuki}
\affil{Department of Earth and Space Science, Graduate School of Science, Osaka University, Toyonaka, Osaka 560-0043, Japan}

\author{Ian A. Bond}
\affil{Institute of Information and Mathematical Sciences, Massey University, Private Bag 102-904, North Shore Mail Centre, Auckland, New Zealand}

\begin{abstract}
Gravitational microlensing is currently the only technique that helps study the Galactic distribution of planets as a function of distance from the Galactic center. 
The Galactic location of a lens system can be uniquely determined only when at least two of the three quantities that determine the mass--distance relations are measured.
However, even if only one mass--distance relation can be obtained, a large sample of microlensing events can be used to statistically discuss the Galactic distribution of the lenses.
In this study, we extract the Galactic distribution of planetary systems from the distribution of the lens-source proper motion,
$\mu_{\rm rel}$, for a given Einstein radius crossing time, $t_{\rm E}$, measured for the 28 planetary events in the statistical sample by \citet{suz16}.
Because microlensing is randomly caused by stars in our Galaxy, the observational distribution can be predicted using a Galactic model.
We incorporate the planet-hosting probability, $P_{\rm host} \propto M_{\rm L}^m R_{\rm L}^r$, into a Galactic model for random-selected stars, where 
$M_{\rm L}$ is the lens mass ($\sim$ host mass), and $R_{\rm L}$ is the Galactocentric distance.
By comparing the observed distribution with the model-predicted $\mu_{\rm rel}$ distribution for a given $t_{\rm E}$ at various combinations of $(m ,r)$,
we obtain an estimate $r = 0.2 \pm 0.4$ under a plausible uniform prior for $m$ of $0<m<2$. 
This indicates that the dependence of the planet frequency on the Galactocentric distance is not large, 
and suggests that the Galactic bulge does have planets.
\end{abstract}


\section{Introduction}
Although more than 3000 planetary systems have been discovered to date, most reside at distances $< 1~{\rm kpc}$ from the Sun\footnote{\dataset[NASA Exoplanet Archive]{https://exoplanetarchive.ipac.caltech.edu/} \citep{ake13}}.
In this regard, gravitational microlensing is a unique technique because it is sensitive to planetary systems over a wide range of distances in our Galaxy, from 
the Galactic disk to the Galactic bulge \citep{gau12}.
Microlensing is currently the only technique capable of investigating the Galactic distribution of planets.
However, no study measuring the distribution has yet been reported, mainly owing to the difficulty in distance measurement.

There are four physical quantities involved in each microlensing event: the lens mass $M_{\rm L}$, distance to the lens $D_{\rm L}$, distance to the source $D_{\rm S}$, 
and lens-source relative proper motion $\mu_{\rm rel}$—--which is given by $\mu_{\rm rel} = |\mu_{\rm L} - \mu_{\rm S}|$, where $\mu_{\rm L}$ is the lens proper motion, and $\mu_{\rm S}$ is the source proper motion.
For most planetary events, two parameters related to these quantities can be commonly measured via light curve modeling, i.e., the Einstein radius crossing time, 
\begin{eqnarray}
t_{\rm E} = \frac{\theta_{\rm E} }{ \mu_{\rm rel}}, 
\end{eqnarray}
and lens-source relative proper motion, $\mu_{\rm rel}$.
Here, $\theta_{\rm E}$ is the angular Einstein radius given by 
\begin{eqnarray}
\theta_{\rm E} = \sqrt{\kappa M_{\rm L} \pi_{\rm rel}}
\end{eqnarray}
where
$\kappa = 8.144 ~{\rm mas} ~M_\odot ^{-1}$ and $\pi_{\rm rel} = {\rm 1\,AU}(D_{\rm L}^{-1} - D_{\rm S}^{-1} )$.

Measuring either the microlens parallax or lens brightness, besides $t_{\rm E}$ and $\mu_{\rm rel}$, is required to determine $M_{\rm L}$ and $D_{\rm L}$,
even assuming that the source star is located in the bulge (i.e., $D_{\rm S} \sim 8~$kpc).
This requirement adds to the complexity of statistical studies focusing on determining the Galactic distribution of planets based on measured distances to the lens.
The microlens parallax is measured only when the lens is relatively close to the Sun \citep{ben10}.
Because of the bias toward closer lenses and the vulnerability to systematic errors \citep{pen16, kos20}, much effort is required to prepare a clean statistical sample of a sufficient number of microlens parallax measurements.

Lens brightness measurements seem to be more robust in terms of susceptibility to systematic errors.
However, to resolve the lens from the source, high-angular-resolution follow-up observations with adaptive optics or observations by the {\it Hubble Space Telescope} ({\it HST})
are required several years after the event.
Thus, a certain amount of time is necessary to obtain a statistical sample of the lens brightness measurements.
A systematic follow-up program for measuring the lens brightness for past planetary events is ongoing using the Keck telescope and {\it HST} \citep{ben18, bha18}, 
and a statistical sample of past planetary events with measurements of $t_{\rm E}$, $\mu_{\rm rel}$, and the lens brightness will be available in the near future.

Meanwhile, this study aims to extract information about the Galactic distribution of planets from a set of measurements of 
the Einstein radius crossing time, $t_{\rm E}$, and the lens-source relative proper motion $\mu_{\rm rel}$ for planetary events.
Thus far, \citet{pen16} attempted to compare the distance distribution of published microlensing planetary systems with predictions based on a Galactic model and 
proposed a possibility that the Galactic bulge might be devoid of planets.
However, their results were affected by an inhomogeneous sample and incorrect microlens parallax measurements \citep{han16}.

In this letter, we compare the $\mu_{\rm rel}$ distribution for a given $t_{\rm E}$ of 28 planetary events from the statistical sample by \citet{suz16} with 
the distribution calculated using a Galactic model optimized for use in microlensing studies \citep{kos21, kos21b}.
We consider a power-law distribution, i.e., $P_{\rm host} \propto M_{\rm L}^m R_{\rm L}^r$, for the planet-hosting probability for lens stars, where $R_{\rm L}$ is 
the Galactocentric distance at which the lens is located.
The comparison of the data and model for various $(m, r)$ values enables us to estimate for the first time
the dependence of the planet-hosting probability on the Galactocentric distance as $P_{\rm host} \propto R_{\rm L}^{0.2 \pm 0.4}$ when a uniform prior for $m$ in $0 < m < 2$ is applied. 

\section{Method}
This work aims to estimate the dependence of the planet-hosting probability on the Galactic location by comparing the $\mu_{\rm rel}$ distribution observed in 
planetary microlensing events with that predicted by a Galactic model.
Although the microlens parallax and/or lens brightness have already been measured for some of the events in the sample,
we here focus on the $t_{\rm E}$ and $\mu_{\rm rel}$ distributions to avoid any bias caused by including them.
The $\mu_{\rm rel}$ distribution alone has no and little information about the lens mass and distance, respectively \citep{pen16}.
However, when combined with $t_{\rm E}$, $\mu_{\rm rel}$ is equivalent to $\theta_{\rm E}$, which yields a mass--distance relation and 
allows us to extract mass and distance information.

\citet{kos20} showed that an observed $\mu_{\rm rel}$ distribution can be compared with a model-expected distribution without considering a detection efficiency correction
once $t_{\rm E}$ is fixed, i.e., 
\begin{eqnarray}
f_{\rm obs} (\mu_{\rm rel} \, | \,  t_{\rm E}) \propto  \Gamma_{\rm Gal} (\mu_{\rm rel} \, | \,  t_{\rm E}) \label{eq-fobs},
\end{eqnarray}
where $f_{\rm obs} (\mu_{\rm rel} \, | \,  t_{\rm E})$ and $\Gamma_{\rm Gal} (\mu_{\rm rel} \, | \,  t_{\rm E})$ are observed and expected $\mu_{\rm rel}$ distributions for given $t_{\rm E}$, respectively.
$\Gamma_{\rm Gal} (\mu_{\rm rel} \, | \,  t_{\rm E}) = \Gamma_{\rm Gal} (\mu_{\rm rel} , t_{\rm E}) / \Gamma_{\rm Gal} (t_{\rm E})$, and $\Gamma_{\rm Gal} ({\bm x})$ is the microlensing event rate
of events with parameter ${\bm x}$, which is calculated using the Galactic model explained in Section \ref{sec-model}.
The observed distribution is given by the sample described in Section \ref{sec-data}.

A key idea behind Eq. (\ref{eq-fobs}) is that the detection efficiency of a microlensing event depends on $t_{\rm E}$ but not on $\mu_{\rm rel}$; thus,
detection efficiency calculations are not necessary to compare observations with simulations for a given $t_{\rm E}$.
Technically, this is not true because the detection of a planetary signal in a light curve depends on the 
source radius crossing time, $t_* = \theta_*/\mu_{\rm rel}$, where $\theta_*$ is the angular source radius.
However, this is expected to have little effect on our results because the dependence of the planet detection efficiency on $t_*$ is 
negligibly small for a mass-ratio of $q \simgt 10^{-4}$ \citep{suz16}, which dominates our sample.

\section{Planetary Microlensing Event Sample} \label{sec-data}
We used 28 planetary microlensing events, which combine 22 planetary events detected by the MOA-II microlensing survey from 2007 to 2012 \citep{suz16} and  
6 planetary events from \citet{gou10} and \citet{cas12}.
This is the same sample as the combined MOA+$\mu$FUN+PLANET sample in \citet{suz16}, except for OGLE-2011-BLG-0950 \citep{cho12}.
The event was excluded from our sample because both recent high-angular resolution imaging observations by Terry et al. (in preparation) and the model-based prior probability calculation by \citet{kos21} suggest that OGLE-2011-BLG-0950 was likely to be a stellar binary-lens event.

There are two events that have degenerate solutions with different $\mu_{\rm rel}$ values in the sample. 
MOA-2011-BLG-262 \citep{ben14} has the fast solution with $\mu_{\rm rel} = 19.6 \pm 1.6$ mas/yr and the slow solution with $\mu_{\rm rel} = 11.6 \pm 0.9$ mas/yr.
We use only the slow solution in our analysis because it has a much larger prior probability as discussed in \citet{ben14}.
MOA-2010-BLG-328 \citep{fur13} has two parallax solutions ($u_0 < 0$ and $u_0 > 0$) and one xallarap solution, and the $\mu_{\rm rel}$ values are measured as $5.71 \pm 0.70~{\rm mas/yr}$, $4.72 \pm 0.79~{\rm mas/yr}$, 
and $4.03 \pm 0.26~{\rm mas/yr}$, respectively.
In the likelihood calculation described in Section \ref{sec-fit}, we equally combine the two parallax solutions, and then equally combine the combined parallax solution with the xallarap solution, i.e., we use $\sqrt{ {\cal L}_{\rm xalla} \sqrt{{\cal L}_{\rm para+} \, {\cal L}_{\rm para-}}}$ as the likelihood for this event.

\begin{figure*}
\centering
\includegraphics[width=15cm]{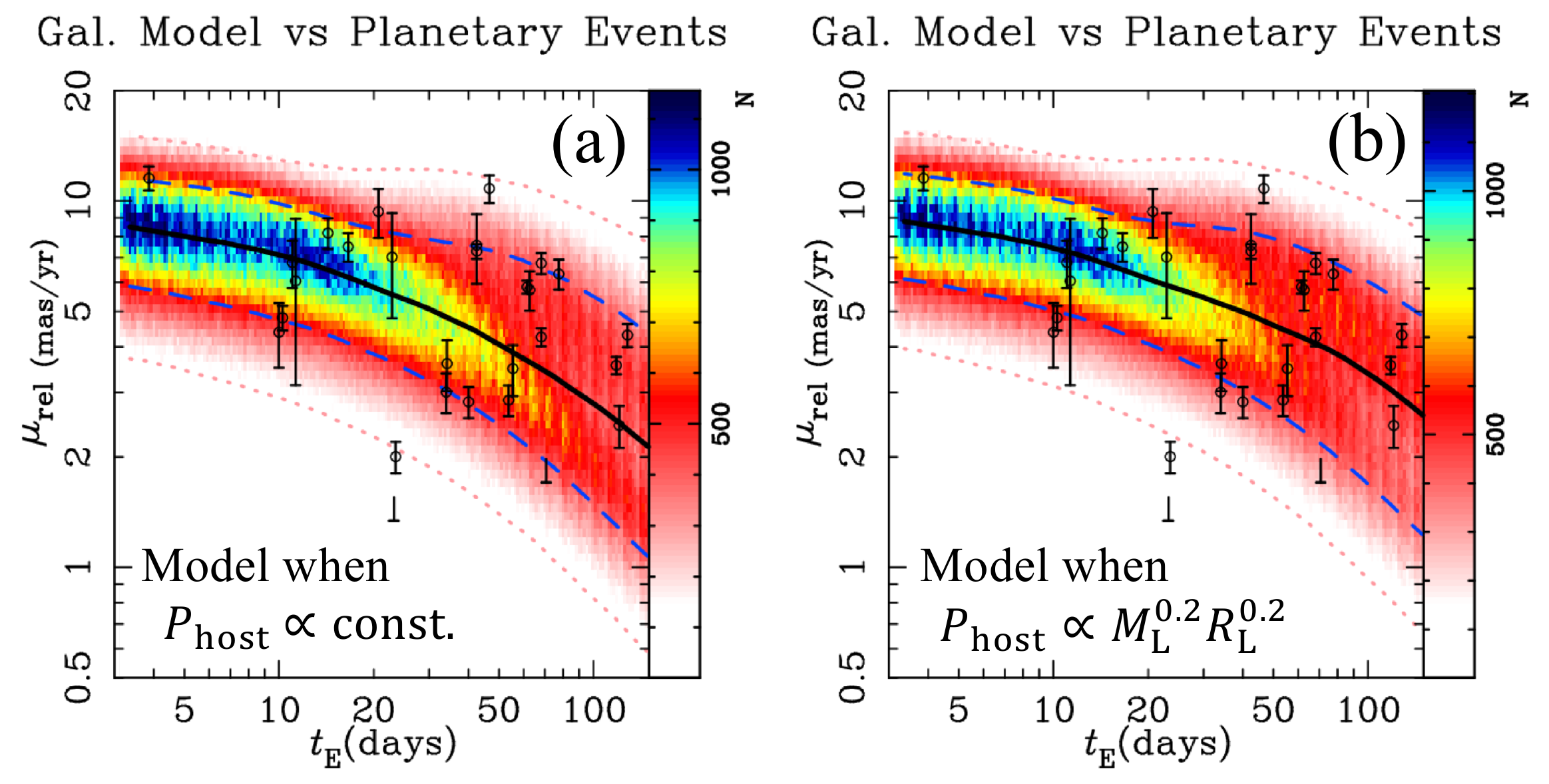}
\caption{Comparison of the $\mu_{\rm rel}$ values of the 28 planetary events from the \citet{suz16} combined sample and Galactic model.
The black open circles show the data, while the color map shows the model-calculated $\mu_{\rm rel}$ distribution for a given $t_{\rm E}$, $\Gamma_{\rm Gal} (\mu_{\rm rel} \, | \,  t_{\rm E})$, 
as a function of fixed $t_{\rm E}$.
The black solid, blue dashed, and magenta dotted lines indicate the median, $1~\sigma$, and $2~\sigma$ for $\Gamma_{\rm Gal} (\mu_{\rm rel} \, | \,  t_{\rm E})$, respectively.
(a) Model with $P_{\rm host} \propto {\rm const}$, i.e., all stars are equally likely to host planets independent of their masses or locations in our Galaxy.
(b) Model with $P_{\rm host} \propto M_{\rm L}^{0.2} R_{\rm L}^{0.2}$, corresponding to the best-fit grid that gives the maximum likelihood.
}
\label{fig-tE_murel}
\end{figure*}

The black open circles in Fig. \ref{fig-tE_murel} represent the $t_{\rm E}$ and $\mu_{\rm rel}$ values of the 28 planetary events; two of them, MOA-2007-BLG-192 \citep{ben08} and 
MOA-2011-BLG-322 \citep{shv14}, have only lower limit measurements on $\mu_{\rm rel}$.
A more detailed description of Fig. \ref{fig-tE_murel} is given in Section \ref{sec-fit}.

\section{Galactic Model for Planet-hosting Stars} \label{sec-model}
To calculate the model of the $\mu_{\rm rel}$ distribution for a given $t_{\rm E}$, $\Gamma_{\rm Gal} (\mu_{\rm rel} \, | \,  t_{\rm E})$, 
we need a Galactic model for planet-hosting stars. 
This combines a model of planet-hosting probability, including the fit parameters investigated in this work, and a Galactic model for random-selected stars, which refers to 
a combination of the stellar mass function, stellar density distribution, and stellar velocity distribution in our Galaxy.

We consider that the planet-hosting probability for the lens depends on the lens mass $M_{\rm L}$ and Galactocentric distance $R_{\rm L}$ at which the lens is located, where
Galactocentric distance refers to the radius in the cylindrical coordinate system with the Galactic center at the origin.
The exact relation remains unknown; thus, in this study, a power law was adopted,
\begin{eqnarray}
P_{\rm host} (M_{\rm L}, R_{\rm L}) \propto M_{\rm L}^m R_{\rm L}^r,  \label{eq-Phost1}
\end{eqnarray}
where $m$ and $r$ are the fit parameters for the powers of $M_{\rm L}$ and $R_{\rm L}$, respectively.
Note that we used $P_{\rm host} (R_{\rm L}) = {\rm const.}$ when $R_{\rm L} < 50~{\rm pc}$ to avoid singularity with $r < 0$.
In the Appendix, we consider a dichotomous model for $P_{\rm host} (R_{\rm L})$ that does not have this singularity.

\subsection{Galactic Model for Random-selected Stars} \label{sec-gal}
\citet{kos21} developed a parametric Galactic model by fitting to Gaia DR2 velocity data \citep{kat18}, 
OGLE-III red clump star count data \citep{nat13}, VIRAC proper motion data \citep{smi18, cla19}, BRAVA radial velocity data \citep{ric07, kun12}, and OGLE-IV star count and microlensing rate data \citep{mro17, mro19}.
All the data except for the Gaia DR2 data correspond to a bulge region of the sky, and the models are optimized for use in microlensing studies toward the Galactic bulge.

There are four versions of models developed in \citet{kos21}, which are denoted by E, G, E+E$_{\rm X}$, and G+G$_{\rm X}$.
Each model consists of a multicomponent thin disk, thick disk, and barred bulge.
For the bulge density profile, exponential functions are used in the E and E+E$_{\rm X}$ models, whereas Gaussian functions are used in the G and G+G$_{\rm X}$ models.
The E and G models have a single-component bulge density profile, whereas the E+E$_{\rm X}$ and G+G$_{\rm X}$ models have an additional component to represent the X-shape structure in the bulge \citep{mcw10, nat10}.
The different bulge profiles used in the four models led to different best-fit parameters for the stellar density, velocity, and initial mass function for all four models.

In this study, we used the E+E$_{\rm X}$ model as our fiducial Galactic model for random-selected stars because it was the most consistent with the data used in \citet{kos21}.
Nevertheless, we also used the other three models to evaluate the systematic errors in our measurement of the planet-hosting probability dependence.

\subsection{Possibilities of Planets around White Dwarfs or Close-binaries} \label{sec-WDCB}
Our sample includes at least one circumbinary planet \citep[OGLE-2007-BLG-349,][]{ben16}.
Because the number of events with lens brightness measurements in our sample is limited,
there is a non-negligible chance that lens objects of other events may also consist of circumbinary planets or even planets around white dwarf hosts.
Thus, both possibilities should be considered in the calculation of the event rate $\Gamma_{\rm Gal} (\mu_{\rm rel} \, | \,  t_{\rm E})$.

We introduce two parameters to account for planet frequencies relative to single stars: $f_{\rm WD}$ for planets around white dwarfs and 
$f_{\rm CB}$ for planets around tight close binaries. Values of $f_{\rm WD} = 1$ or $f_{\rm CB} = 1$ indicate the same planet frequency as for a single star with the same mass; if $f_{\rm WD} = 0$ or $f_{\rm CB} = 0$, there is zero probability of hosting planets.
Determining the details of the planet abundance around white dwarfs or close binary systems is beyond the scope of this study;
thus, we adopted ($f_{\rm WD}$, $f_{\rm CB}$) = (1, 1) for our fiducial model and used cases ($f_{\rm WD}$, $f_{\rm CB}$) = (0, 1), (1, 0), and (0, 0) to evaluate the systematic errors in our $r$ estimate.

To consider the white dwarf population and tight close-binary population in the model, we followed \citet{kos21}, where 
the initial-final mass relation of $M_{\rm WD} = 0.109 M_{\rm ini} + 0.394 M_{\odot}$ \citep{kal08} is used for the white dwarfs. Further, 
the binary distribution developed by \citet{kos20b}, together with a detection threshold based on a combination of the central caustic size of 
a hypothetical binary and the event impact parameter $u_0$, are used for tight close-binaries.

We assume that the planet-hosting probability for a neutron star or black hole is zero.
The assumption is based on the study by \citet{beh20} that found no planets in an 11-year dataset for 45 pulsars.
If this rareness of planets around neutron stars is attributed to the progenitor's explosion (i.e., supernova), 
planets are also likely to be rare around black holes.

\section{Maximum Likelihood Analysis} \label{sec-fit}
The color maps in Fig. \ref{fig-tE_murel} are two examples of the model-calculated $\mu_{\rm rel}$ distribution for 
given $t_{\rm E}$, $\Gamma_{\rm Gal} (\mu_{\rm rel} \, | \,  t_{\rm E})$, $(m, r) = (0, 0)$ in Fig. \ref{fig-tE_murel}(a) and 
$(m, r) = (0.2, 0.2)$ in Fig. \ref{fig-tE_murel}(b).
The distributions are calculated over the range $0.50 < \log (t_{\rm E}/{\rm days}) < 2.20$. First, this range is divided into 34 bins
of width 0.05 dex each. Then, $10^5$ artificial events are generated using our fiducial Galactic model for planet-hosting stars, i.e., 
the E+E$_{\rm X}$ model with ($f_{\rm WD}$, $f_{\rm CB}$) = (1, 1).
We selected a Galactic coordinate of $(l, b) = (1.0^{\circ}, -2.2^{\circ})$ for these plots.
The coordinates of each event are used in the relative likelihood calculations presented below.

Fig.~\ref{fig-tE_murel}~(a) shows that the observed $\mu_{\rm rel}$ distribution is already in good agreement with the model, in which 
no planet-hosting probability dependence on the stellar mass or position in our Galaxy is assumed.
This means that $(m, r) = (0, 0)$ is an acceptable planet-hosting probability dependence under the current constraint imposed by the data.
Below, we calculate the relative likelihood for various combinations of $(m, r)$ and obtain the probability distributions for these parameters.

\subsection{Definition of Likelihood}
We define the likelihood for a combination of $m$ and $r$ as
\begin{eqnarray}
{\cal L} (m, r) = \prod_{i} \Gamma_{\rm Gal} (\mu_{{\rm rel}, i} \, | \,  t_{{\rm E}, i} \, ; \, m, r),  \label{eq-L}
\end{eqnarray}
where $\mu_{{\rm rel}, i}$ and $t_{{\rm E}, i}$ are the ones observed for $i$th event, and
 $\Gamma_{\rm Gal} (\mu_{{\rm rel}, i} \, | \,  t_{{\rm E}, i} \, ; \, m, r)$ is the model-calculated probability of $\mu_{{\rm rel}, i}$ for 
given $t_{{\rm E}, i}$ when the planet-hosting probability is $P_{\rm host} \propto  M_{\rm L}^m R_{\rm L}^r$. 
The product is taken over all 28 events in our sample.

For each event $i$, in the calculation of $\Gamma_{\rm Gal} (\mu_{{\rm rel}, i} \, | \,  t_{{\rm E}, i} \, ; \, m, r)$, 
the corresponding parameters are used, e.g., the Galactic coordinate and impact parameter $u_0$ affecting the close-binary possibilities.

\subsection{Results} \label{sec-res}

\begin{figure}
\centering
\includegraphics[width=8.8cm]{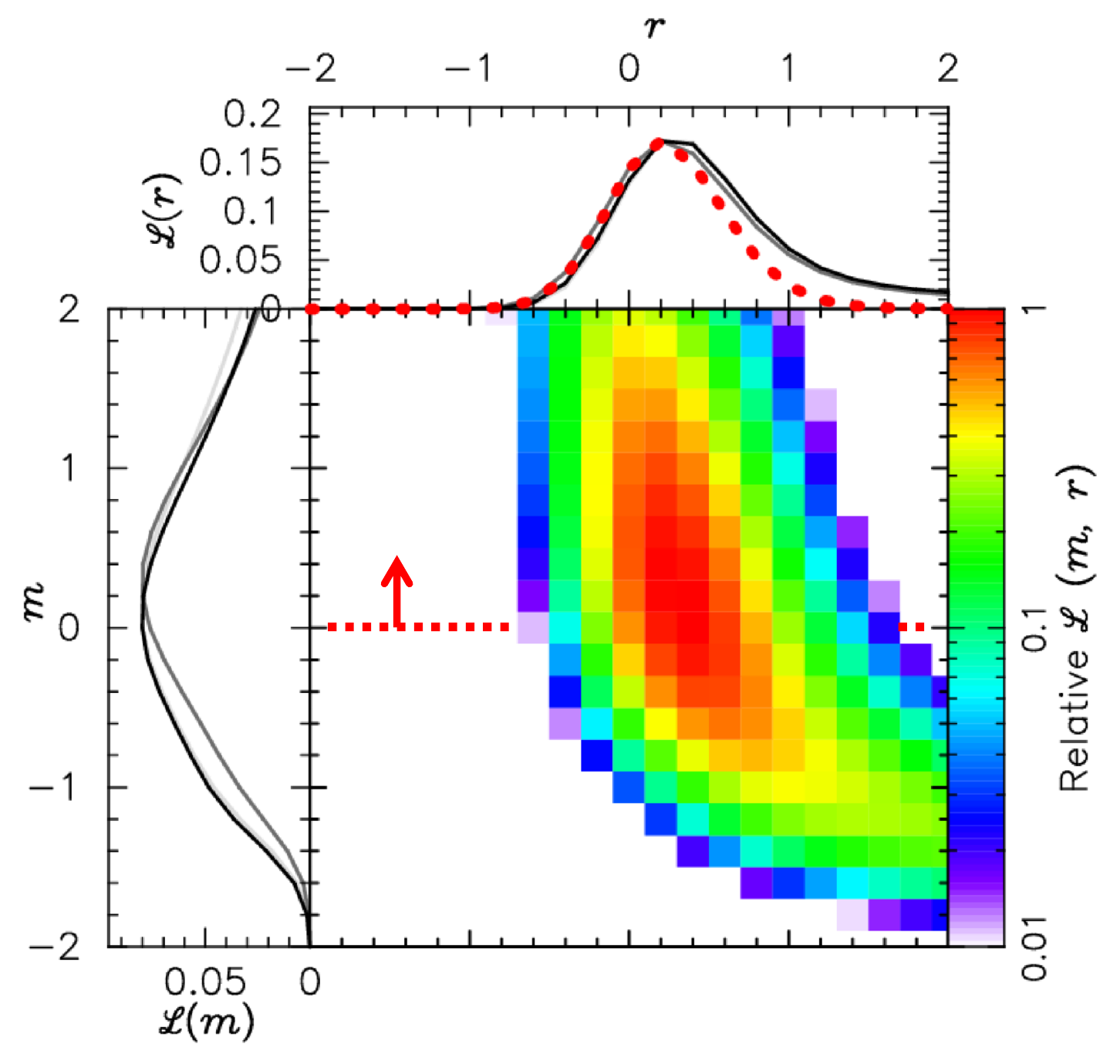}
\caption{Relative likelihood distribution as a function of $m$ and $r$ for the E+E$_{\rm X}$ fiducial Galactic model with ($f_{\rm WD}$, $f_{\rm CB}$) = (1, 1).
In the top panel, probability distributions of $r$ are shown in solid black and dotted red, integrated ${\cal L} (m, r)$ over $-2 < m < 2$ and $0 < m < 2$, respectively.
The dark and light gray lines in the edge panels are the marginalized probability distributions when 
($f_{\rm WD}$, $f_{\rm CB}$) = (0, 1) and ($f_{\rm WD}$, $f_{\rm CB}$) = (1, 0), respectively. 
(Note that the light gray line almost overlaps with the black line.)
}
\label{fig-L}
\end{figure}

We calculated the likelihood given by Eq. (\ref{eq-L}) over the ranges of $-2 < m < 2$ and $-2 < r < 2$ at intervals of 0.2, for both $m$ and $r$.
This corresponds to applying uniform prior distributions for the two parameters between -2 and 2, which reflects that 
there are moderate fractions of planets in all categories of low-mass lenses \citep[e.g., MOA-2008-BLG-310,][]{jan10, bha17, kos20b}, high-mass lens \citep[e.g., OGLE-2012-BLG-0026,][]{han13, bea16},
distant lens \citep[e.g., MOA-2007-BLG-400,][]{don09, bha20}, and close lens \citep[e.g., OGLE-2006-BLG-109,][]{gau08, ben10}.
The planet frequency dependence on $M_{\rm L}$ or $R_{\rm L}$ would not be as extreme as $|m| > 2$ or $|r| > 2$.

Fig. \ref{fig-L} shows the results, where a clear negative correlation between $m$ and $r$ is seen.
The shape of the correlation is attributed to the mass--distance relation given by the angular Einstein radius $\theta_{\rm E}$, a parameter equivalent to $\mu_{\rm rel}$ for a given $t_{\rm E}$.
Because certain $\theta_{\rm E}$ values can be explained by both a low-mass lens close to the Sun and a high-mass lens close to the source, the 
data distribution can be reproduced by both small $m$ with large $r$ and large $m$ with small $r$.
This degeneracy can be disentangled by including observational information of either the microlens parallax, lens brightness, or Einstein radius crossing time $t_{\rm E}$ with a
proper detection efficiency correction, which all are interesting future works.

On the other hand, the location of the negatively correlated distribution on the $r$ vs. $m$ plane is determined by the data and Galactic model. Fig. \ref{fig-L} shows a high likelihood value at $(m, r) = (0, 0)$, as expected from Fig.~\ref{fig-tE_murel}~(a).
For comparison, we show the $\Gamma_{\rm Gal} (\mu_{\rm rel} \, | \, t_{\rm E})$ distribution using the best-fit grid value of $(m , r) = (0.2, 0.2)$ in Fig.~\ref{fig-tE_murel}~(b).
The correlation between $r$ and $m$ becomes weaker when $m$ increases, and a stronger constraint can be placed on $r$ than $m$, for which
we have estimated $r = 0.4^{+0.6}_{-0.4}$.

Because it is difficult to put a useful constraint on the $m$ value based on the current data, we considered a stronger uniform prior on $m$ in the range $0 < m < 2$.
This is a plausible prior considering the ongoing high-angular resolution follow-up program on past events \citep{ben18}, which revealed that two events (out of seven published) 
have lens host stars that are much more massive than those expected by Bayesian analysis using a Galactic model for random-selected stars \citep{van20, bha20}.\footnote{
Our preliminary result from a statistical study on the \citet{suz16} sample based on mass measurements implies $m > 0$ with a $3~\sigma$ confidence level, as reported by DPB in the
Exoplanet Demographics Conference in November 2020. The talk can be found on \dataset[https://www.youtube.com/watch?v=5TpOKjHSS10]{https://www.youtube.com/watch?v=5TpOKjHSS10}.}
For giant planets, the possibility of $m > 0$ is also supported by other techniques. 
\citet{joh10} estimated $m = 1.0 \pm 0.3$ in $\simlt 2.5~$AU from a radial velocity survey, and  \citet{nie19} estimated $m = 2.0 \pm 1.0$ 
in 10--100~AU from a direct imaging survey.

As shown by the red dotted curve in the top panel of Fig.~\ref{fig-L}, applying the narrower prior of  $0 < m < 2$ can almost entirely exclude the possibility of $r > 1$.
The median and $1~\sigma$ error is $r = 0.2 \pm 0.4$ for the fiducial E+E$_{\rm X}$ Galactic model with ($f_{\rm WD}$, $f_{\rm CB}$) = (1, 1).

Finally, to evaluate the systematic error due to model selection, we repeated this analysis using the E, G, and G+G$_{\rm X}$ models and 
for ($f_{\rm WD}$, $f_{\rm CB}$) = (1, 0), (0, 1), and (0, 0).
We found that the median value of $r$ becomes 0.3--0.5 and 0.2--0.3 for the uniform priors in $-2 < m < 2$ and $0 < m < 2$, respectively. 
Thus, the systematic errors seem to be much smaller than the statistical errors.
Table \ref{tab-results} summarizes the results of our likelihood analysis.

\begin{deluxetable}{ccccccccccccc}
\tablecaption{Results of the maximum likelihood analysis. \label{tab-results}}
\tablehead{
\colhead{Range of uniform prior}       &                              \multicolumn{2}{c}{$r$}                           &   &    \colhead{$m$ \tablenotemark{c}} \\
                             & \colhead{Fiducial \tablenotemark{a}} & \colhead{Sys. range \tablenotemark{b}}  &   &   \colhead{Fiducial \tablenotemark{a}}
}                                                                                                                                           
\startdata                   
     $-2 < m < 2$  &   $0.4^{+0.6}_{-0.4}$  &    0.3--0.5         &   &  $0.2 \pm 1.0$  \\
 \ \, $0 < m < 2$  &   $0.2 \pm 0.4$        &    0.2--0.3         &   &  $0.7^{+0.8}_{-0.6}$ \\
\enddata 
\tablenotetext{a}{Median and $1~\sigma$ error for the E+E$_{\rm X}$ model and $(f_{\rm WD}, f_{\rm CB}) = (1, 1)$.}
\tablenotetext{b}{Variation of median values when other Galactic models or other combinations of $(f_{\rm WD}, f_{\rm CB})$ are applied.}
\tablenotetext{c}{Listed for completeness. The estimates of $m$ are dominated by the prior applied.} 
\end{deluxetable}

\section{Discussion and Conclusion} \label{sec-dis}
We estimated the planet-hosting probability dependence of $P_{\rm host} \propto R_{\rm L}^{0.2 \pm 0.4}$ under a uniform prior distribution of $m$ in $0 < m < 2$, which 
suggests no large dependence of the planet frequency on the Galactocentric distance.
The bulge region has a very different stellar environment from the solar neighborhood, including stellar densities $\simgt 10$ times higher and 
an older, alpha-enhanced stellar population.
Observations of the solar neighborhood have shown that there is a correlation between stellar metallicity and the occurrence of giant planets \citep{fis05, joh10}, 
and have also suggested that close encounters with other stars may affect the evolution of planetary systems \citep{win20}.
Therefore, due to the abovementioned environmental differences, the planet frequency in the bulge may differ significantly from that in the solar neighborhood.
Although our results are still inconclusive, they might imply that cold planets orbiting beyond the H$_2$O snow line also commonly exist 
in the bulge region regardless of such differences.

Because all exoplanet detection techniques have sensitivities that complement each other, it is important to combine different techniques for a comprehensive understanding of 
planet formation and evolution processes \citep{gau20}.
One of the largest uncertainties when comparing the exoplanet population discovered via other exoplanet detection methods with microlensing planets is a possible large dependence of the planet frequency on Galactic location.
Our results show that such a dependence is not very large, and one might be able to compare them without considering the difference in Galactic location.
A small dependence of the planet frequency on the Galactic location is also supported by \citet{suz16}. \citet{suz16} showed that the planet frequency from the MOA-II microlensing survey---which is an averaged value for stars in the galactic disk and bulge---is consistent with the frequency of cold gas giants from radial velocity studies \citep{bon13, mon14}.

On the other hand, the full consistency with $r = 0$ contrasts with the analysis by \citet{pen16}, who found a small $p$-value of 5.0 $\times 10^{-4}$ for 
a model in which the planetary frequency of the bulge is the same as that of the disk.
The difference is most likely due to the contamination of excessively close lenses with incorrect parallax measurements in their sample.
\citet{pen16} themselves discussed this possibility, and questioned the results of several events with large microlens parallax values. 
In fact, among the questioned events, OGLE-2013-BLG-0723 was a stellar binary event \citep{han16}.
The large microlens parallax claimed for MOA-2007-BLG-192 \citep{ben08} has recently been shown to disappear when the data are detrended 
for color-dependent differential refraction \citep{ben12}, although this has not yet been published.
Nevertheless, our results are still consistent with the \citet{pen16}'s estimate of the bulge-to-disk ratio of 
planet frequency, $f_{\rm bulge} < 0.54$, as shown in the Appendix.

Our estimate of $r = 0.2 \pm 0.4$ still has a moderately large uncertainty, and there could be a non-negligible dependence of 
planet frequency on the Galactocentric distance $R_{\rm L}$.
In particular, although $m > 0$ seems to be thus far plausible \citep{van20, bha20}, if the possibility of $m < 0$ is considered, the uncertainty and preference for disk planets could increase.
This is owing to the negative correlation between $m$ and $r$, as shown in Fig. \ref{fig-L}, which is attributed to our use of only 
one parameter (i.e., $\theta_{\rm E}$) that provides a lens mass--distance relation in the analysis.
Thus, further constraints should emerge by including constraints from other mass--distance relations, which can be provided by a statistical sample of either microlens parallax or lens brightness measurements.

\acknowledgments
NK was supported by the JSPS overseas research fellowship.
DPB and NK were supported by NASA through grant NASA-80NSSC18K0274 and award number 80GSFC17M0002.
DS was supported by JSPS KAKENHI Grant Number JP19KK082 and JP20H04754.

\appendix

\section{Dichotomous Model for Planet-hosting Probability} \label{sec-fDB}
One might think that the power law model, $P_{\rm host} \propto R_{\rm L}^r$, is not physical because it diverges at $R_{\rm L} = 0$ when $r < 0$.
Here, we consider a dichotomous model for the dependence of the planet-hosting probability on the lens Galactic location,
\begin{equation}
P_{\rm host} (R_{\rm L}) =
\begin{cases}
P_{\rm host, D} & \text{ when $R_{\rm L} > 2 {\rm kpc}$} \\
P_{\rm host, B} & \text{ when $R_{\rm L} \leq 2 {\rm kpc}$,} \, \label{eq-Phost2}
\end{cases}
\end{equation}
and we use the ratio between the two constants, $f_{\rm B/D} \equiv P_{\rm host, B} / P_{\rm host, D}$, as the fit parameter instead of $r$.
This is a more directly comparable model to that used by \citet{pen16}, who considered the bulge-to-disk ratio of planet frequency 
in the \citet{han03} Galactic model, $f_{\rm bulge}$.
Note that there is still a subtle difference between the two parameters---$f_{\rm B/D} = 0$ indicates that there are no planets at $R_{\rm L} \leq 2 {\rm kpc}$,  
whereas $f_{\rm bulge} = 0$ allows planet-hosting stars in the disk component to exist at $R_{\rm L} \leq 2 {\rm kpc}$.

\begin{figure}
\centering
\includegraphics[width=8.8cm]{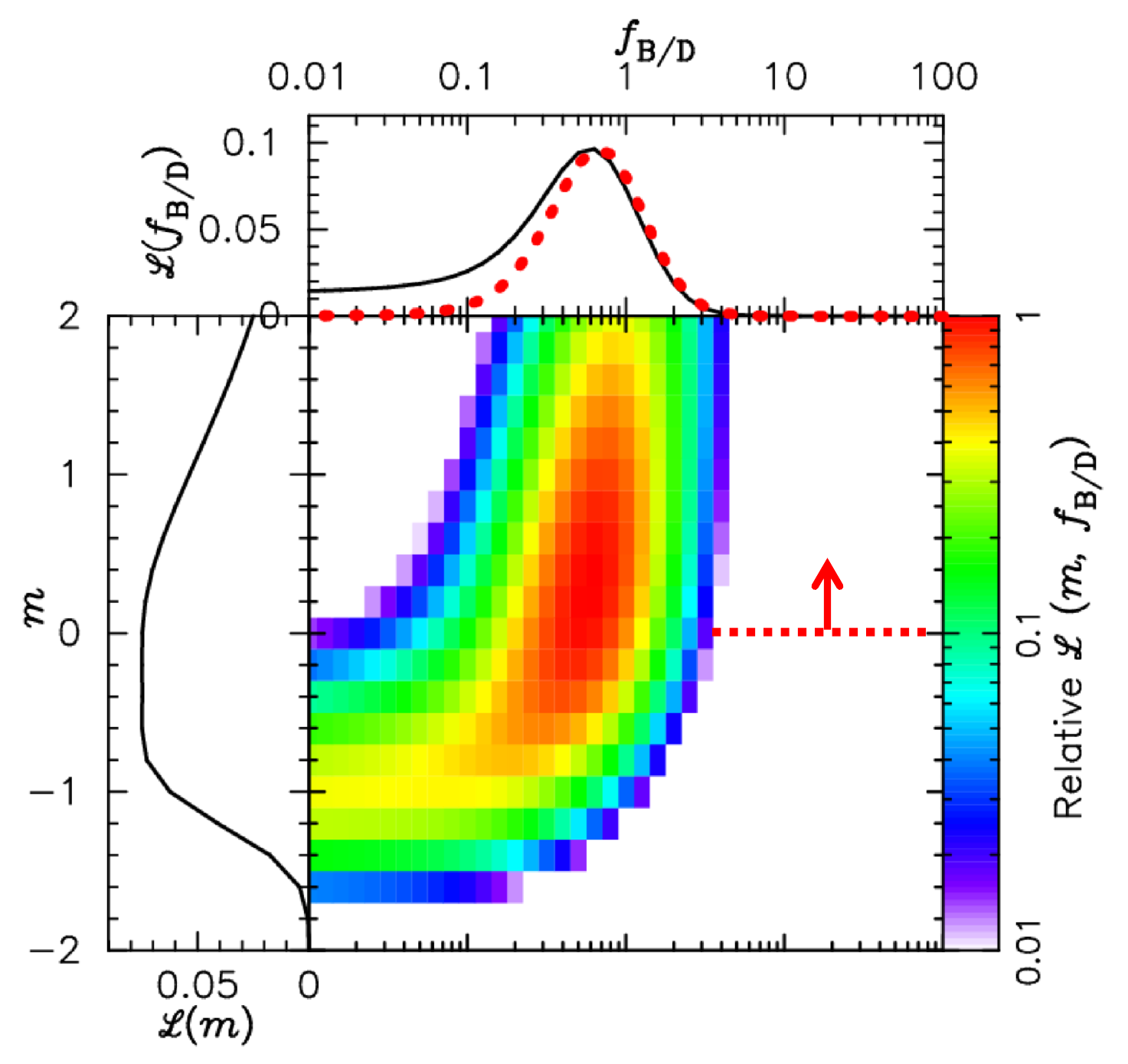}
\caption{Same as Fig. \ref{fig-L}, but for $f_{\rm B/D}$ instead of $r$.}
\label{fig-L2}
\end{figure}

\begin{deluxetable}{ccccccccccccc}
\tablecaption{$f_{\rm B/D}$ estimated from the maximum likelihood analysis.} \label{tab-results2}
\tablehead{
\colhead{Range of uniform prior}       &                         \multicolumn{2}{c}{$f_{\rm B/D}$}                        \\
                                       & \colhead{Fiducial \tablenotemark{a}} & \colhead{Sys. range \tablenotemark{b}}
}
\startdata
     $-2 < m < 2$                      &  $0.48^{+0.59}_{-0.33}$              &    0.42 -- 0.52                           \\
 \ \, $0 < m < 2$                      &  $0.63^{+0.61}_{-0.33}$              &    0.56 -- 0.65                           \\
\enddata
\tablenotetext{a}{Median and $1~\sigma$ error with the E+E$_{\rm X}$ model and $(f_{\rm WD}, f_{\rm CB}) = (1, 1)$.}
\tablenotetext{b}{Variation of median values when other Galactic models or other combinations of $(f_{\rm WD}, f_{\rm CB})$ are applied.}
\end{deluxetable}

Fig. \ref{fig-L2} and Table \ref{tab-results2} show the results of the maximum likelihood analysis using $f_{\rm B/D}$ instead of $r$, where 
the uniform prior for $\log f_{\rm B/D}$ in $-2 < \log f_{\rm B/D} < 2$ is assumed.
With the uniform prior for $m$ in $0 < m < 2$, we estimate $f_{\rm B/D} = 0.63^{+0.61}_{-0.33}$, which is consistent with both $f_{\rm B/D} = 1$ and
$f_{\rm bulge} < 0.54$; the upper limit estimate for $f_{\rm bulge}$ was taken from a $p$-value threshold of 0.01 by \citet{pen16}.

\end{document}